\documentclass[aps,prx,twocolumn,superscriptaddress,longbibliography]{revtex4-2}
\usepackage{graphicx}
\usepackage{multirow}
\usepackage{makecell}
\usepackage{booktabs}
\usepackage{lineno}
\usepackage{amsmath}
\usepackage[table]{xcolor}
\usepackage{physics}

\begin{document}
    \title{Robust Orbital-Selective Flat Bands in Transition-Metal Oxychlorides}

    \author{Xiangyu Luo$^{1,\dag,*}$, Ludovica Zullo$^{2,\dag,*}$, Sahaj Patel$^{1,\dag}$, Dongjin Oh$^{1}$, Qian Song$^{1}$, Asish K. Kundu$^{3}$, Anil Rajapitamahuni$^{3}$,  Elio Vescovo$^{3}$, Natalia Olszowska$^{4}$, Rafał Kurleto$^{4}$, Dawid Wutke$^{4}$, Giorgio Sangiovanni$^{2}$ and Riccardo Comin$^{1,*}$\\ \small\textit{$^{1}$Department of Physics, Massachusetts Institute of Technology, Cambridge, MA 02139, USA
    \\$^{2}$Institut f\"{u}r Theoretische Physik und Astrophysik and W\"{u}rzburg-Dresden Cluster of Excellence ct.qmat, Universit\"{a}t W\"{u}rzburg, 97074 W\"{u}rzburg, Germany
    \\$^{3}$National Synchrotron Light Source II, Brookhaven National Laboratory, Upton, NY, USA
    \\$^{4}$Solaris National Synchrotron Radiation Centre, Jagiellonian University, Czerwone Maki 98, 30-392 Kraków, Poland
    \\$^{\dag}$These authors contributed equally to this work.
    \\$^{*}$Corresponding author: xyluo@mit.edu, ludovica.zullo@uni-wuerzburg.de and rcomin@mit.edu}}

    \date{\today}

    \begin{abstract}
    Flat electronic bands, which amplify electron correlations by quenching kinetic energy, provide an ideal foundation for exotic quantum phases. However, prevailing strategies---including geometrically frustrated lattices, moiré superlattices and heavy-fermion physics---suffer from inherent trade-offs among robustness, tunability and orbital selectivity, limiting their broad applicability. Here, we unveil an intrinsic orbital-selective flat-band mechanism in the van der Waals materials NbOCl$_2$ and TaOCl$_2$, directly observed by angle-resolved photoemission spectroscopy (ARPES) and understood through density functional theory (DFT) and Wannier analysis. Crucially, we experimentally demonstrate that this momentum-independent flat band exhibits remarkable robustness, surviving from the bulk crystal down to the few-layer limit at room temperature. Our theoretical analysis traces its origin to the hybridization between Nb-$d_{z^{2}}$ orbital chains and the Lieb-like $d_{x^{2}-y^{2}}$ sublattice, which is further reinforced by Peierls dimerization. Our findings not only establish transition-metal oxychlorides as a robust and tunable platform for flat-band–driven correlated phases under ambient conditions, but also uncover a new orbital-selective design principle for realizing flat bands in quantum materials.
    \end{abstract}

	\maketitle

    Flat electronic bands (FBs) have emerged as a central motif in quantum materials research, providing a unifying platform where strong electron–electron correlations intertwine with a nontrivial topology\cite{ETang2011,TNeupert2011,KSun2011,ZLiu2014,DLeykam2018,JXYin2022,NRegnault2022,JGCheckelsky2024}. Within these FBs, the suppression of electronic kinetic energy amplifies Coulomb interactions, so that even moderate on-site repulsion can drive the system far from conventional Fermi-liquid behavior \cite{SPaschen2021}. Indeed, the dominance of the on-site Coulomb interaction $U$ over electron kinetic energy $t$ offers a universal mechanism for correlation effects, giving rise to unconventional superconductivity\cite{SPeotta2015,AJulku2016,YCao2018,LBalents2020,EYAndrei2020,BROrtiz2020}, Mott phase\cite{MImada1998,TOhashi2006,YCao2018_2,YDWang2020,SGao2023}, non-Fermi-liquid transport\cite{JHuang2024,LYe2024} and topologically ordered phases hosting integer or fractional excitations, as first realized in Landau levels \cite{LLandau1930,KVKlitzing1980,RBLaughlin1983} and more recently in moiré materials \cite{EJBergholtz2013,TDevakul2021,YXie2021,JCai2023,YZeng2023,HPark2023}. Known routes to realizing FBs can be broadly grouped into three categories: (i) geometrically frustrated lattices, where quantum destructive interference suppresses hopping \cite{BSutherland1986,EHLieb1989,AMielke1991,AMielke1992,PMNeves2024}, as experimentally observed in kagome \cite{ZLiu2020,MKang2020,MKang2020_2,LYe2024,BSong2025}, Lieb\cite{ADevarakonda2025} and pyrochlore \cite{JPWakefield2023,JHuang2024} lattices; (ii) moiré superlattices, which flatten bands via momentum-space band folding\cite{RBistritzer2011,KPNuckolls2024,MIBUtama2021,SLisi2021}; and (iii) correlated heavy-fermion systems, where orbital localization narrows the bandwidth\cite{GRStewart1984,QSi2010,QChen2017,VAPosey2024}. Despite their distinct microscopic origins, these approaches share the essential feature of quenching kinetic energy to amplify electron correlations.

    However, all established strategies face intrinsic trade-offs between robustness, tunability, and orbital selectivity. Moiré FBs engineered via momentum-space folding as well as quantum destructive interference induced FBs are vulnerable to broadening from interlayer coupling or disorder, often requiring cryogenic stabilization\cite{EYAndrei2020,DMKennes2021,NRegnault2022,PMNeves2024}. Correlation-driven FBs in heavy-fermion systems, though profoundly localized, suffer from the extreme atomic localization of $f$ orbitals, which severely limits fine tunability\cite{QSi2010,SPaschen2021,JGCheckelsky2024}; furthermore, the dual nature of $f$ orbitals—partially localized and strongly hybridized with conduction d electrons—blurs the distinction between localized and itinerant states. Symmetry-protected FBs demand preservation of delicate crystalline symmetries that can be fragile to disorder or strain in realistic settings\cite{DLeykam2018,ARamachandran2017}. The trilemma of achieving strong orbital selectivity, room-temperature stability, and multi-parameter control has hindered the development of tunable FB platforms for correlated-electron studies. Consequently, an ideal FB material that simultaneously fulfills these requirements has yet to be realized.

    In this work, we resolve this challenge by identifying an intrinsic orbital-selective FB mechanism in the layered van der Waals material niobium and tantalum oxychlorides (NbOCl$_{2}$ and TaOCl$_{2}$). Their crystal structure—featuring Peierls-dimerized chains and Lieb-like lattice—selectively quenches kinetic energy in Nb/Ta $d_{z^{2}}$ orbitals through bond-length disproportionation and electron localization. This yields a momentum-independent isolated FB that coexists with dispersive states, as directly observed by angle-resolved photoemission spectroscopy (ARPES) in both compounds. Owing to their intrinsic two-dimensional nature and weak interlayer coupling, the FB persists from bulk crystals down to few-layer flakes and remains stable at room temperature without the need for moiré engineering. Density functional theory (DFT) calculations attribute its origin to the hybridization between the quasi-1D Nb $d_{z^{2}}$ chain and the Lieb-like O ($p_x$)–Nb ($d_{x^{2}-y^{2}}$)-Cl ($p_y$) network which is further reinforced by Peierls distortion. This orbital selectivity confines electrons to the flat band, highlighting their localized nature, while orbital geometry provides an additional degree of freedom that can facilitate flat-band formation and offers a platform to tune correlation strength via the orbital degree of freedom\cite{YZheng2025}. By combining orbital selectivity, room-temperature stability and multi-parameter tunability within an intrinsic material platform, NbOCl$_{2}$ and TaOCl$_{2}$ establish a new paradigm for engineering correlated quantum matter beyond the constraints of existing FB systems.

    Figure 1a shows that NbOCl$_{2}$ crystallizes in the monoclinic space group C$_{2}$ (Supplementary Fig. S1), forming a quasi-2D framework with weak van der Waals stacking along the a-axis\cite{JRijnsdorp1978,HHillebrecht1997,LYe2023,QGuo2023}. Within each layer, edge-sharing NbOCl$_{2}$ distorted octahedra form chains along b-axis through O bridges and connect along c-axis via Cl bridges (Fig. 1b), giving rise to a strong in-plane anisotropy. The lattice exhibits two cooperative structural distortions: along c direction, Peierls distortion dimerizes the Nb chains into alternating short ($d_1$) and long ($d_2$) Nb–Nb bonds (Fig. 1c); along b direction, Nb ions undergo a collective off-center displacement toward one bridging oxygen, producing alternating Nb–O bond lengths (Fig. 1b). These coupled distortions break inversion symmetry, stabilize a low-energy polar configuration, and generate a robust in-plane ferroelectric polarization at room temperature\cite{CLiu2023}, accompanied by pronounced second-harmonic generation and other optical responses in such system\cite{LYe2023,QGuo2023,IAbdelwahab2022,QGuo2024}. Owing to the exceptionally weak interlayer coupling, large-area monolayer and few-layer flakes can be readily obtained by mechanical exfoliation and retain their structural and electronic integrity at room temperature over extended periods, making them ideal for surface-sensitive probes\cite{QGuo2023}.

    Examining the lattice geometry projected onto the b–c plane reveals a network closely resembling a Lieb lattice. Although the chlorine atoms in NbOCl$_{2}$ are slightly displaced from the atomic plane, their bonding environment allows the system to be effectively modeled as Lieb-like lattice (Fig. 1b), capturing the essential symmetry and hopping pathways. Fig. 1d shows the contrast between the kagome lattice and the Lieb lattice, where corner-sharing triangles in the kagome lattices yield FBs at the band edges, whereas Lieb lattices host them near the spectrum center in idealized tight-binding models. For Nb$^{4+}$ (4$d^{1}$), the effects of the crystal field lower the $d_{z^{2}}$ orbital relative to the other $d$ orbitals, making it the main contributor to the states near $E_{F}$ (Fig. 1e)\cite{LYe2023}. In the absence of lattice distortion, the configuration of the $d_{z^{2}}$ electrons would produce a simple metallic state with one itinerant electron in the orbital $d_{z^{2}}$. However, Peierls dimerization along the Nb-Cl-Nb chains splits the $d_{z^{2}}$ manifold into bonding (BB) and antibonding (AB) states, thereby opening a gap and driving a metal–insulator transition. The corresponding schematic density of states (DOS) shows a fully occupied BB and an empty AB, consistent with the insulating ground state\cite{QGuo2023}. This orbital–lattice reconstruction provides a natural multi-component route to kinetic-energy quenching, as further corroborated by DFT and Wannier-function analysis in the following. Fig. 1f presents the DFT band structure for dimerized monolayer NbOCl$_{2}$, with the rectangular Brillouin zone (BZ) defined by $k_{x}$ along the $\Gamma$–X direction (parallel to the $b$-axis) and $k_{y}$ along the $\Gamma$–Y direction (parallel to the $c$-axis). The calculations reveal a narrow, well-isolated FB derived predominantly from $d_{z^{2}}$ orbitals near the Fermi level over entire BZ, with $E_{F}$ lying inside the gap between the FB and higher conduction bands, consistent with previous theoretical work\cite{MAMohebpour2024}. The persistence of this FB in the ultrathin limit underscores its intrinsic nature. Fig. 1g summarizes the cooperative mechanism: Peierls dimerization suppresses inter-dimer $d_{z^{2}}$ hopping and enhances orbital localization, while Lieb-like connectivity enforces destructive interference that also quenches dispersion. Acting together, these factors stabilize an orbital-selective flat band that is both structurally and electronically robust, providing a natural platform for exploring correlation-driven phases.

    To directly probe the electronic structure and verify the predicted orbital-selective flat band, ARPES measurements were performed on high-quality NbOCl$_{2}$ single crystals at room temperature to avoid charging effects. Fig. 2a shows constant energy contours acquired with linear horizontal (LH) polarized light at h$\nu$ = 66\,eV, spanning binding energies from $E_{B}$ = 2.2 to 7.5\,eV. The pronounced anisotropy in spectral weight along the $\Gamma$–X and $\Gamma$–Y directions reflects the material’s low crystallographic symmetry. High-symmetry points (blue circles) and momentum cuts (Cut 1 and Cut 2, grey lines) were selected for detailed band dispersion analysis. Fig. 2b presents the band dispersion along $\Gamma$–X direction (Cut 1), measured at h$\nu$ = 100\,eV. A remarkably narrow, nearly dispersionless feature appears near $E_{B}$ $\sim$ 2.2\,eV, denoted as the FB. Energy distribution curves (EDCs) integrated at $\Gamma$ and X points show clear peaks at this energy, confirming the FB’s momentum-independent character. DFT calculations for monolayer NbOCl$_{2}$ along the same path (right panel) reproduce the FB alongside more dispersive states at higher binding energies, in excellent agreement with experimental observations. Fig. 2c shows the momentum cut along $\Gamma$–Y direction (Cut 2) and the corresponding EDCs at $\Gamma$ and Y point. The FB retains its extreme flatness, while the valence bands below exhibit reduced dispersion and narrower bandwidth compared to those along $\Gamma$–X direction, indicating anisotropic conductivity\cite{CLiu2023,YFang2021}, likely originating from the anisotropic Cl- and O-mediated hopping pathways and the different electronegativity between Cl and O atoms. Corresponding DFT calculations (Fig. 2c, right panel) also confirm the flatness feature for the $d_{z^{2}}$ dominated FB along $\Gamma$–Y direction, underscoring its isotropic localization in the whole momentum space. To further clarify the orbital character of the FB, Fig. 2d shows the orbital-resolved projected density of states (PDOS) for monolayer NbOCl$_2$. The FB near the VBM is dominated by Nb $d_{z^{2}}$ orbitals, with a finite admixture of $d_{x^{2}-y^{2}}$ and Cl/O $p$ states, because the low-symmetry environment breaks the strict orthogonality between $d_{z^{2}}$ and $d_{x^{2}-y^{2}}$ orbitals, allowing Cl-mediated hybridization that imparts a finite $d_{x^{2}-y^{2}}$ component near $E_{F}$. The inset displays the real-space charge density of the FB's wavefunction from DFT, highlighting their strong  confinement around Nb sites and the toroidal in-plane character of the $d_{z^{2}}$ orbital. Photon-energy-dependent measurements (Fig. 2e) trace the $\Gamma$–X dispersion over a wide $k_{z}$ range (40–140 eV). While the FB intensity is stronger at lower photon energies, its binding energy remains fixed with negligible $k_{z}$ dependence, confirming its intrinsic two-dimensional nature in NbOCl$_{2}$. In contrast, the spectral weight of some deeper valence bands varies with photon energy due to matrix element effects. Additional photon-energy-dependent constant energy contours and momentum cuts showing the invariance of the flat-band binding energy across multiple kz values are provided in Supplementary Fig. S2. Remarkably, these ARPES measurements reveal that the flat band remains entirely dispersionless along all three momentum directions ($k_x$, $k_y$, and $k_z$). Furthermore, temperature-dependent ARPES and related EDC analyses (Supplementary Fig. S3) indicate that the flat-band energy and spectral weight do not show significant shifts across the studied temperature range (including across the reported FE–AFE crossover\cite{MHuang2024}), thereby supporting its room-temperature robustness.

    To assess the universality of this orbital-selective FB mechanism, we synthesized the 5$d$ analogue TaOCl$_{2}$\cite{NNg2024} and performed comparative ARPES measurements on both compounds. Fig. 3a directly contrasts the electronic structures of NbOCl$_{2}$ and TaOCl$_{2}$ along the high-symmetry paths in the first BZ. In both materials, ARPES spectra acquired at h$\nu$ = 70\,eV combine momentum cuts obtained with linear horizontal (LH) and vertical (LV) polarizations. The experimental results are in overall agreement with the corresponding monolayer DFT band calculations. Momentum-independent FBs are clearly resolved across the entire BZ. Integrated photoemission spectra over a broad binding-energy range (Fig. 3b) show well-defined core-level peaks from Nb/Ta, O and Cl in each compound. Slight core-level shifts relative to the elemental references indicate variations in oxidation state and chemical environment. Fig. 3c shows polarization-dependent spectra along the $\Gamma$–X direction in both compounds. The FB intensity is maximized under LH polarization but strongly suppressed under LV. Along $\Gamma$–Y direction (Fig. 3d), both compounds exhibit a nearly dispersionless FB near $\Gamma$ with similar polarization dependence, while the spectral weight decreases toward and beyond Y point. The full set of EDC stacks and Lorentzian fits of the peak positions across $\Gamma$–X and $\Gamma$–Y show negligible peak shifts (Supplementary Fig. S4). The extracted bandwidth is below 70\,meV, which is consistent with the previous theoretical predictions ($\sim$80\,meV)\cite{MAMohebpour2024}. Notably, the FB in TaOCl$_{2}$ displays a slightly larger bandwidth compared to NbOCl$_{2}$. The extremely narrow bandwidth of the flat band in NbOCl$_2$, arising from orbital frustration, gives rise to a pronounced peak in the density of states (Fig. 2d and Supplementary Fig. S5) and highlights the potential role of strong electron correlations. EDCs extracted at $\Gamma$, X, and Y (Figs. 3(e,f)) further confirm that FB peaks are strongest at $\Gamma$ and further enhanced at X under LH polarization, but are strongly suppressed under LV; along $\Gamma$–Y direction, the FB is intense at $\Gamma$ but weakens at Y under LH and nearly vanishes under LV. Extended polarization-dependent ARPES data on TaOCl$_{2}$ (Supplementary Figs. S6 and S7) show the same LH/LV intensity contrast. These polarization trends indicate that the FB is primarily derived from the even-parity $d_{z^{2}}$ orbital. Because the $d_{z^{2}}$ state is nearly rotationally symmetric in the $x$-$y$ plane, comparable intensity at X and Y would be expected if the $d_{z^{2}}$ orbital were the only contributor. The enhanced spectral weight of the FB at X, alongside its suppression at Y, indicates a minor admixture of $d_{x^{2}-y^{2}}$ orbital with anisotropic in-plane lobes, which redistributes the matrix elements. This anisotropic spectral-weight evolution supports the presence of a small $d_{x^{2}-y^{2}}$ component hybridized with the dominant $d_{z^{2}}$ character, as discussed in Fig. 2d and Supplementary Fig. S5. Comparing the two systems, the FB position in TaOCl$_{2}$ is slightly closer to $E_F$ than in NbOCl$_{2}$, indicating a smaller band gap ($\sim$2.05\,eV) and suggesting a potentially higher electrical conductivity. In the valence bands below the FB, TaOCl$_{2}$ exhibits a broader bandwidth and larger interband splitting, as indicated by the black dashed lines (Fig. 3e). These features are consistent with the more extended nature of 5$d$ orbitals, which enhance both interdimer and interlayer coupling, thereby slightly relaxing the kinetic-energy quenching. Equivalent EDC analyses along $\Gamma$–Y direction (Fig. 3f) yield the same trends. Taken together, these comparative results demonstrate that the orbital-selective FB mechanism is robust across the Nb/Ta series, while the degree of flatness and the band position can be tuned via orbital extent. This provides a practical materials-design knob for engineering correlation strength and bandwidth in quantum materials.

    It has been theoretically predicted that the electronic properties of NbOCl$_{2}$ remain nearly unchanged from the three-dimensional bulk down to the two-dimensional limit--- an exceptionally rare behavior among layered materials\cite{YJia2019,BMortazavi2022,PHelmer2025}. Having confirmed the universality of the orbital-selective FB across the Nb/Ta series in bulk samples, we next experimentally examine its persistence in the few-layer regime. Fig. 4 shows that the FB in NbOCl$_{2}$ is remarkably robust upon thickness reduction. We fabricated graphene/NbOCl$_{2}$/hBN heterostructures in which mechanically exfoliated few-layer NbOCl$_{2}$ flakes were encapsulated by graphene and supported on hBN with patterned Au contacts on a SiO$_{2}$/Si substrate, enabling high-quality micro-ARPES measurements (Fig. 4a). The optical image (Fig. 4b) reveals the morphology of the stacking heterostructure, while atomic force microscopy (AFM) confirms a representative flake thickness of $\sim$2\,nm (three layers; Fig. 4c). ARPES spectra along $\Gamma$–X direction for the few-layer device (Fig. 4d) reveal a sharp, dispersionless feature at $\sim$2\,eV below the Fermi level, as pointed out by the orange arrow. For comparison, graphene alone (Fig. 4e) has been measured to show no such FB feature and confirm its origin from NbOCl$_{2}$. Bulk NbOCl$_{2}$ samples measured under identical conditions (Fig. 4f) exhibits an almost identical FB. EDCs at the $\Gamma$ point (Fig. 4g) further highlight the close similarity between few-layer (purple) and bulk (red) spectra, with only minimal energy shifts. This negligible change from bulk to few-layer suggests that the exceptional properties of NbOCl$_{2}$, including its strong optical responses that are likely preserved even in the monolayer limit\cite{QGuo2023}. While our data conclusively demonstrate the persistence of the flat band in few-layer flakes, its stability in the strict monolayer limit is strongly suggested by the extreme 2D nature of the electronic structure observed in bulk (Fig. 2e) and the excellent agreement with monolayer DFT calculations. The negligible evolution from bulk to trilayer further supports the notion that the monolayer will host the same robust flat band. In practical terms, this means that devices based on relatively easy-to-fabricate few-layer flakes can already deliver performance comparable to that of ideal monolayers, substantially lowering the barrier for applications.

    Having established the good agreement between ARPES measurements and DFT calculations (Figs. 2 and 3), we now turn to reveal the microscopic origin of the FB. Earlier theoretical studies highlighted two essential ingredients: (i) orbital mixing between Nb and Cl atoms, and (ii) the critical role of Peierls distortion in driving the band-gap opening and emergent chirality\cite{BMortazavi2022,CLiu2023,MAMohebpour2024, MGAmigo2025}. Here, we unify and extend these concepts to develop a comprehensive framework for flat-band formation in NbOCl$_{2}$. First, to explore the influence of the dimerization process, we compare a 2$\times$1 Peierls-distorted supercell in real space with a 2$\times$1 non-distorted supercell generated by doubling the unit cell along the $y$-axis---the direction of the charge-density-wave (CDW) modulation. This construction allows a direct comparison between distorted and undistorted phases. We further investigate the evolution of lattice distortion by gradually shifting atomic coordinates of each atom between the non-CDW (A) and CDW (B) phases, following the relation $F$ = $x$A + $y$B with $x$ + $y$ = 1 as defined in Ref.\cite{XLYu2017}. The two limited cases $x$ = 1, $y$ = 0 and $x$ = 0, $y$ = 1 correspond to the undistorted and fully distorted structures, respectively, while intermediate values represent virtual snapshots of the dimerization process (Supplementary Fig. S8). Following the progression of the Peierls distortion, we can map the gap-opening mechanism. This interpolation reveals that even a modest distortion ($x$ = 0.75, $y$ = 0.25) breaks the $-y/y$ mirror symmetry in real space and induces a small gap along the Y–S path of the BZ between two bands near the Fermi level (Supplementary Fig. S8b). From this point onward, the gap-opening process proceeds inexorably, culminating in the fully developed CDW phase (B). This analysis establishes that the gap formation is driven predominantly by the emergence of the CDW order. Throughout the A-B evolution, the most pronounced macroscopic consequence is this symmetry-breaking gap opening, whereas the overall topology of the band structure remains largely preserved.

    We next investigate the orbital character of the A and B phases to evaluate the crucial role of the orbital hybridization in driving the FB formation. Figs. 5a and 5b present the orbital-projected band structures for both phases, revealing that the 28 bands above and below the band gap derive primarily from Nb-$d$, Cl-$p$ and O-$p$ states. Focusing on the FB at the valence band maximum (VBM), there is a clear orbital mixing: although the main contribution arises from Nb-$d_{z^2}$ (red) orbitals, some weight also comes from Nb-$d_{x^{2}-y^{2}}$ (blue) and Cl-$p_{z}$ (black), with a minor contribution from Cl-$p_{y}$ (lime). Moreover, closer inspection shows that another set of conduction bands also opens a gap as the system evolves from phase A to phase B. These states are predominantly derived from Nb-$d_{yz}$ orbitals (purple). By comparing these two distinct gap-opening processes, we can gain insight into the formation of the FB. A minimal tight-binding (TB) model is developed using hoppings derived from Wannierization of the niobium $d_{z^{2}}$ and $d_{yz}$ DFT bands. This model includes two Wannier functions, $d_{z^{2}}$ ($d_{yz}$), centred on each Nb atom, Nb$_{1}$ and Nb$_{2}$, of the 2$\times$1 supercell. The TB Hamiltonian is constructed as:
    \begin{equation}
    \scalebox{0.80}{$
    \begin{pmatrix}
    E_{1} + t_{01}e^{ik_{y}(R + r_{j} - r_{i})} & t_{a}e^{ik_{y} (R + r_{j} - r_{i})} + t_{b}e^{ik_{y} (R + r_{j} - r_{i})} \\
    t_{a}e^{-ik_{y} (R + r_{j} - r_{i})} + t_{b}e^{-ik_{y} (R + r_{j} - r_{i})} & E_{2} + t_{02} e^{ik_{y} (R + r_{j} - r_{i})}
    \end{pmatrix}$}
    \end{equation}
    where $\mathbf{r}_i$ and $\mathbf{r}_j$ are the atomic coordinates and $\mathbf{R}$ is a lattice translation. $E_1$ ($E_2$) denotes the on-site energy for atom 1 (2). The parameter $t_{01}$ ($t_{02}$) describes the inter-cell hopping between atom 1 (2) and its copy in the next cell. In the off-diagonal terms, $t_a$ and $t_b$ represents the intra-cell interactions between atom 1 and atom 2. Applying this Hamiltonian to the non-CDW scenario reveals that $t_{01} = t_{02}$ and $t_a$ = $t_b$ due to symmetry, resulting in a closed gap for both orbital types. In contrast, in the CDW case, $t_{01}$ = $t_{02}$ since the two atoms host the same type of orbitals, but $t_a < t_b$ due to dimerization. The computed band structures are shown in the small panels in Fig. 5(a,b). This analysis maps the complex multi-orbital problem onto an Su–Schrieffer–Heeger (SSH)-like Hamiltonian, where the gap-opening mechanism originates from the Peierls distortion. Crucially, the orbital geometry of the $d_{yz}$ state, confined to the $y$–$z$ plane, prevents electron hopping along the $x$ direction, thereby realizing an effective dimensionality reduction. In this picture, these two bands with extreme flatness may be viewed as a quasi-one-dimensional SSH chain of Nb-$d_{yz}$ and $d_{z^2}$ orbitals mediated by Cl atoms, whose band structure is gapped out and flattened due to symmetry breaking. According to the SSH model, the band gap magnitude is 2$\times$$|t_a - t_b|$ for both orbital types. The Nb $d_{z^2}$ and $d_{yz}$ orbitals exhibit substantially different band gaps of 2.31\,eV and 0.37\,eV, respectively. This pronounced disparity calls for further investigation: indeed, even within this minimal two-band Wannierization, we are still implicitly capturing the interaction of $d_{z^2}$ and $d_{yz}$ with the surrounding orbital environment. As a result, the band gap size difference points to two fundamentally different interactions with the rest of the local chemical environment, which remain to be fully resolved.

    To this end, we performed a full Wannierization, including all the Nb-$d$, Cl-$p$ and O-$p$ orbitals and extracted all the most meaningful hopping parameters between atoms in the supercell. Our Wannierization analysis decomposes the complex electronic structure of NbOCl$_{2}$ into three independent orbital-lattice modules (Fig. 5(c-e)). We first classify the Nb $d$ orbitals by symmetry into $t_{2g}$ ($d_{xy}$, $d_{yz}$, $d_{xz}$) and $e_g$ ($d_{z^2}$, $d_{x^2-y^2}$) manifolds. Within the $t_{2g}$ set, the $d_{xy}$ and $d_{xz}$ orbitals each couple to a single O-$p$ and Cl-$p$ orbital, forming two distinct three-site clusters: O($p_y$)-Nb($d_{xy}$)-Cl($p_x$) and O($p_z$)-Nb($d_{xz}$)-Cl($p_x$). Each cluster maps onto a separate Lieb-like sublattice (Fig. 5c). Consistent with previous predictions, the remaining $d_{yz}$ orbital forms an SSH chain along the CDW axis ($y$-axis), mediated by Cl $p_y$ and $p_z$ orbitals, and shows no coupling to oxygen (Fig. 5d and Supplementary Fig. S9). Along the orthogonal $x$ direction, the orientation of the $d_{yz}$ orbital precludes overlap with neighboring oxygen orbitals, while at the same time it fails to hybridize with the other $t_{2g}$ states that possess finite spatial extension along $x$ direction (Supplementary Table S1 and Fig. S9). With both direct and indirect hopping channels blocked, $d_{yz}$ electrons are confined into quasi-one-dimensional chains, effectively realizing an SSH model. This reduction in dimensionality, driven by orbital geometry rather than structural constraints alone, emerges from our DFT analysis of NbOCl$_2$ as a natural extension of the SSH model beyond its conventional one-dimensional setting. In particular, the directional anisotropy of the $d{yz}$ orbital enforces quasi-1D confinement even within a 2D lattice, demonstrating that orbital shape can itself generate low-dimensional electronic behavior in higher-dimensional crystals. This orbital-geometry–driven blocking of hopping not only stabilizes the $d_{yz}$ flat band but also establishes a general principle of effective dimensional reduction. However, this straightforward picture, valid for $t_{2g}$ orbitals, becomes considerably more intricate for the $e_g$ orbitals and plays a pivotal role in understanding the FB at the VBM.

    The Wannierization reveals a significant on-site hybridization between the $e_g$ orbitals, with an order of 1\,eV. The distorted octahedral coordination of Nb enables strong coupling between these $e_g$ states. Furthermore, the toroidal in-plane shape of the Nb-$d_{z^2}$ orbital enables hopping along the $x$-axis with the O-$p_x$ orbitals (Supplementary Table S1). As a result, the underlying Lieb-like lattice formed by the O($p_x$)-Nb($d_{x^2-y^2}$)-Cl($p_y$) triangle becomes intricately intertwined with the quasi-one-dimensional SSH chain formed by Nb-$d_{z^2}$ orbitals via Cl $p_y$ and $p_z$ states (Fig. 5e). The interplay between these two networks gives rise to the FB at the VBM, further reinforced by the Peierls-distortion-induced dimerization. This hopping analysis forms the basis for constructing the tight-binding models for each unique sublattice (Fig. 5(c-e)). Collectively, these orbital–lattice configurations effectively explain the emergence of robust orbital-selective FB and other associated valence-band features revealed by ARPES. To further validate our model through the orbital selections derived by the Wannierization, we performed the Slater-Koster parameterization of each $d$-$p$ interaction. The results summarized in Supplementary Table S1, quantify the interactions of Nb $d$ -O $p$ and Nb $d$ -Cl $p$ and show complete consistency with our tight-binding models, with finite overlaps taking place exclusively in the orbital clusters mentioned above. Our theoretical framework is broadly applicable because it relies on orbital characterization and Slater-Koster overlaps that can extend to the entire family of transition-metal oxydihalides. Beyond its generality, the conceptual novelty of this model paves the way for the realization of flat-band systems via sublattice hybridization and orbital-geometry-driven dimensional reduction, combined with the dimerization process.

    By combining experiment and theory, we have identified and experimentally verified an intrinsic orbital-selective flat band in the van der Waals layered materials NbOCl$_2$ and TaOCl$_2$. This flat band originates from the hybridization of Peierls-dimerized SSH chains with a Lieb-like sublattice. This momentum-independent flat band is robust from bulk to few-layer limit, stable at room temperature and tunable via compositional variations (transition-metal ion or halogen substitution), offering simultaneous orbital selectivity, structural stability and multi-parameter control. These findings position transition-metal oxychlorides as a versatile platform for engineering correlated quantum phases, offering a pathway to flat-band–driven physics without relying on moiré superlattices or extreme conditions. Future work should target the controlled exploration of correlated states (e.g., via strain, gating, or doping), synthesis of related compounds to test universality  and direct probing of emergent orders (e.g., leveraging intrinsic ferroelectricity and inducing magnetism to enable multiferroicity)—unlocking new quantum phenomena at accessible temperatures.

    \vspace{3mm}

    \noindent {\bf Acknowledgement}\\
    This work was supported, in part, by the Air Force Office of Scientific Research (AFOSR) under grant FA9550-22-1-0432 (ARPES, material synthesis and characterization). R.C. gratefully acknowledges support from the Alexander von Humboldt Foundation through the Friedrich Wilhelm Bessel Research Award. This research used resources from the ESM beamline of the National Synchrotron Light Source II, a U.S. Department of Energy (DOE) Office of Science User Facility operated for the DOE Office of Science by Brookhaven National Laboratory under Contract No. DE-SC0012704. The ARPES setup from the URANOS beamline was developed under the provision of the Polish Ministry and Higher Education project Support for research and development with the use of research infra-structure of the National Synchrotron Radiation Centre “SOLARIS” under Contract No. 1/SOL/2021/2. This material is based upon work supported by the National Science Foundation Graduate Research Fellowship Program under Grant No. 2141064. L. Z. and G. S. gratefully acknowledge the Gauss Centre for Supercomputing e.V. (https://www.gauss-centre.eu) for funding this project by providing computing time on the GCS Super-computer SuperMUC-NG at Leibniz Supercomputing Centre (https://www.lrz.de). L. Z. and G. S. gratefully acknowledge for funding support from the Deutsche Forschungsgemeinschaft (DFG, German Research Foundation) under Germany’s Excellence Strategy through the W\"{u}rzburg-Dresden Cluster of Excellence on Complexity and Topology in Quantum Matter ct.qmat (EXC 2147, Project ID 390858490).

    \vspace{3mm}

    \noindent {\bf Author Contributions}\\
     X.L., L.Z., G.S. and R.C. proposed and designed the research. X.L., D.O. and S.P. carried out the ARPES experiments at the National Synchrotron Light Source II and Solaris National Synchrotron Radiation Centre, with support from A.K.K., A.R., E.V., N.O., R.K. and D.W. X.L. analyzed the data. X.L. grew the single crystals and prepared the few-layer samples with the help of S.P. and Q.S. L.Z. and G.S. performed DFT calculations. All authors contributed to writing the paper and participating in discussions and comments on the paper. X.L., L.Z., G.S. and R.C. oversaw the project.

    \vspace{3mm}

    \noindent{\bf Competing interests}\\
     The authors declare no competing interests.

    \vspace{3mm}

    \noindent {\bf Data and code availability}\\
    All data needed to evaluate the conclusions in the paper are available within the article and its Supplementary Information files. All raw data and code generated during the current study are available from the corresponding author upon request.\\

    \vspace{3mm}

    \noindent{\bf Methods}

    \noindent{\bf Sample growth and characterization}\\
    Single crystals of NbOCl$_2$ and TaOCl$_2$ were synthesized via chemical vapour transport. High-purity Nb$_2$O$_5$ (Ta$_2$O$_5$; 99.99\%), Nb (Ta; 99.99\%) and NbCl$_5$ (TaCl$_5$; 99.99\%) powders were mixed in a stoichiometric ratio of 1:1:2, sealed in evacuated quartz tubes and placed in a horizontal three-zone furnace following the procedure in Ref.\cite{QGuo2023,NNg2024}. Rectangular-shaped crystals were obtained with a typical growth duration lasting from three weeks to one month. Phase purity and composition were confirmed by energy-dispersive X-ray spectroscopy (EDS). Polarization-resolved Raman spectra (Supplementary Fig. S1) further corroborate phase purity and provide vibrational fingerprints that validate the crystallographic orientation.\\

    \noindent{\bf Device fabrication}\\
    The device used for micro-ARPES measurements was fabricated by assembling a van der Waals heterostructure consisting of few-layer NbOCl$_2$, graphene, and hBN on a SiO$_2$ (300 nm)/Si substrate with pre-patterned Au electrodes. Few-layer NbOCl$_2$ flakes were mechanically exfoliated inside a glovebox and transferred onto hBN using a polymer-based dry transfer technique. A graphene layer was subsequently stacked on top of NbOCl$_2$ to provide encapsulation and environmental protection. Electrical contacts were defined by depositing Ti (5\,nm)/Au (30\,nm) electrodes on SiO$_2$ (300\,nm)/Si substrate using standard photolithography methods. The entire stacking and transfer procedure was carried out under inert conditions to minimize exposure of NbOCl$_2$ to ambient moisture and oxygen. The resulting graphene/NbOCl$_2$/hBN heterostructures enabled stable and high-quality micro-ARPES measurements. The sample thickness was determined by atomic force microscopy (AFMWorkshop HR) measurements, using a silicon probe in tapping mode.\\

    \noindent{\bf ARPES measurements}\\
    ARPES experiments were carried out at beamlines URANOS of the SOLARIS National Synchrotron Radiation Centre and the National Synchrotron Light Source II ESM beamlines of Brookhaven National Laboratory. Both beamlines were equipped with Scienta DA30 electron analysers. The results were reproduced at both facilities. The angular and total energy resolutions were set to 0.1$^o$ and less than 20\,meV. For the measurements of the few-layer samples at ESM, micro-focused synchrotron radiation was employed with a beam spot $\sim$10$\times$5\,$\mu$$m^{2}$. The flake positions were identified by real space mapping. Bulk crystals were in situ cleaved prior to measurement. Critically, all samples—both bulk and exfoliated few-layer flakes—were measured at room temperature (T = 300\,K) under ultra-high vacuum conditions  (better than 9$\times$10$^{-11}$ Torr). The Fermi level was calibrated using a clean polycrystalline Au reference in electrical contact with the sample.\\

    \noindent{\bf DFT calculations}\\
    First-principles calculations were performed in the framework of density functional theory (DFT), as implemented in the \textsf{Quantum-ESPRESSO} package\cite{Giannozzi2009, Giannozzi2017}.
    Projected augmented wave (PAW) pseudopotentials were selected for all the calculations,  in the generalized gradient approximation (GGA) based on the Perdew-Burke-Ernzerhof (PBE) scheme.
    The plane wave basis set was truncated using a cutoff energy of $50$ Ry for the plane waves and $500$ Ry to represent the charge density in all calculations.
    The Brillouin zone was sampled by $16\times8\times1$ Monkhorst-Pack $k$-point grid.
    Considering the $z$-axis as the stacking direction for the bulk, the supercell size was kept fixed along that direction for all the analyzed monolayer structures.
    A $\sim13$ {\AA} vacuum space was added along the out-of-plane direction, proven to be sufficient to prevent periodic replicas from interacting with each other.
    For all the considered structures, the atomic positions were optimized using the conjugate gradient algorithm in the PBE\cite{JPPerdew1996} scheme, with a convergence criterion such that the total energy difference between consecutive structural optimization steps was less than $10^{-5}$ eV and all components of the forces acting on the atoms must be less than $10^{-4}$ eV/{\AA}. Wannierization was carried out in the Wannier90\cite{GPizzi2020} framework by considering a total of $28$ Wannier-functions including every Nb-$d$, O-$p$ and Cl-$p$ orbitals per formula unit and a Monkhorst-Pack $k$-point grid of $12\times12\times1$ points, proved sufficient to appropriately reproduce the monolayer's ab-initio band structure. Finally, the tight-binding parameters were extracted from the Wannier Hamiltonian by means of the PyTB code\cite{SCoh2022}, which was used to perform tight-binding models for each of the individual sublattice.\\

    \begin{figure*}[h]
    \begin{center}
    \includegraphics[width=2\columnwidth,angle=0]{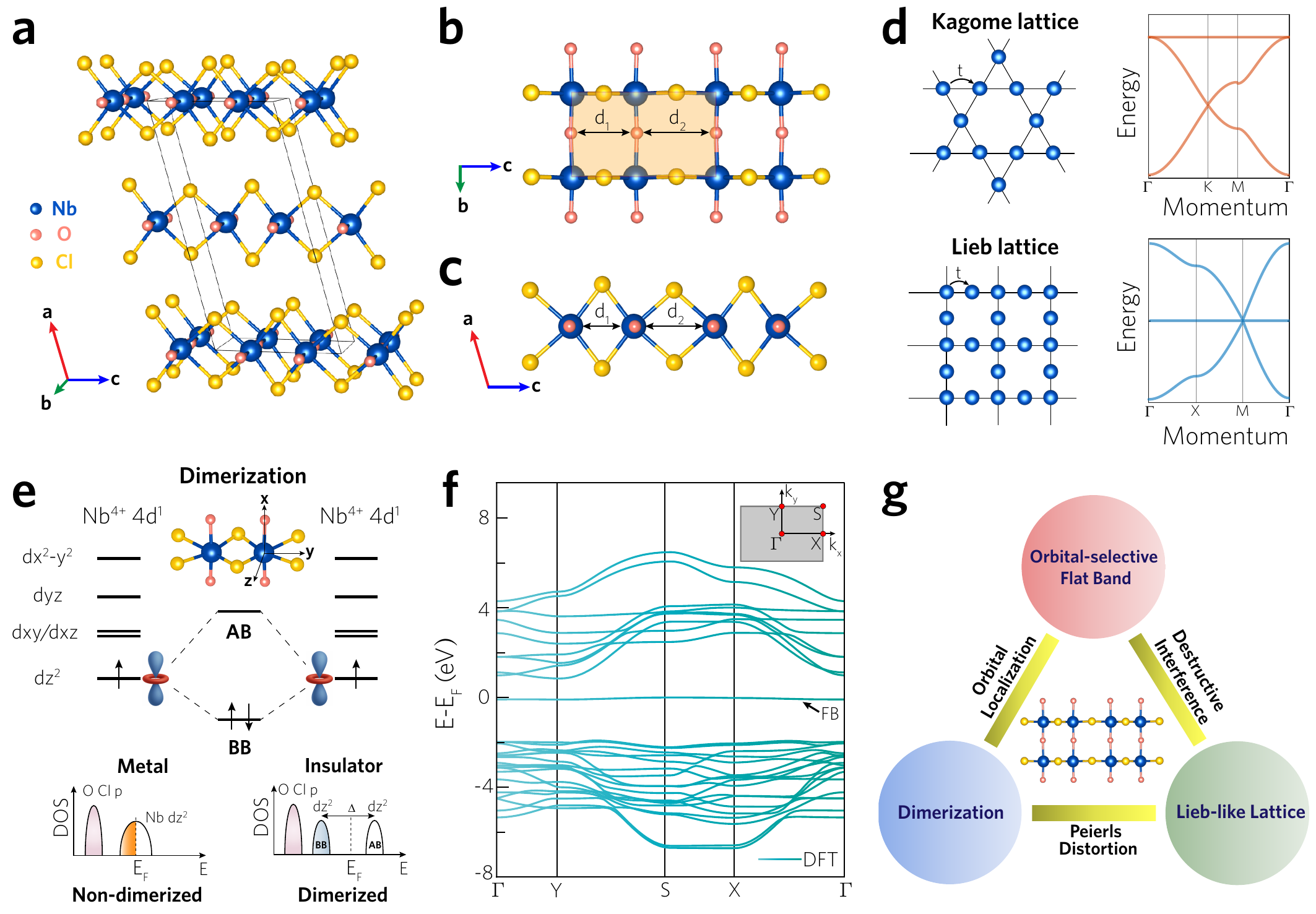}
    \end{center}
    \caption{{\bf Crystal structure, cooperative Peierls distortions and emergence of an orbital-selective flat band in NbOCl$_{2}$.} (a-c) Crystal structure of NbOCl$_{2}$ viewed along various crystallographic directions. Nb (blue), O (red) and Cl (yellow) atoms form quasi-2D layers stacked along the a axis. Within each layer, alternating Nb–Nb bond lengths $d_{1}$ and $d_{2}$ highlight pronounced dimerization. (d) Schematic lattices and tight-binding band structures of kagome and Lieb systems, where symmetry-protected flat bands arise from destructive interference. Owing to distinct lattice geometries, the flat band positions differ between these two systems. NbOCl$_{2}$ adopts a related Lieb-like lattice configuration. (e) Orbital energy levels of Nb$^{4+}$ ($4d^{1}$) atoms and bonding configurations in the dimerized state. (f) Density functional theory (DFT) band structure of the monolayer dimerized NbOCl$_{2}$. The rectangular BZ is defined with $k_{x}$ along the $\Gamma$–X direction (parallel to the b axis) and $k_y$ along the $\Gamma$–Y direction (parallel to the c axis). The narrow, well-isolated flat band (FB) near the Fermi level remains robust in the ultrathin limit, with the Fermi level inside the gap between the FB and dispersive bands. (g) Schematic of the cooperative mechanism: dimerization driven by Peierls distortion, Lieb-like lattice interference and orbital localization collectively stabilize the orbital-selective flat band.}
    \end{figure*}

    \begin{figure*}[tp]
    \begin{center}
    \includegraphics[width=2\columnwidth,angle=0]{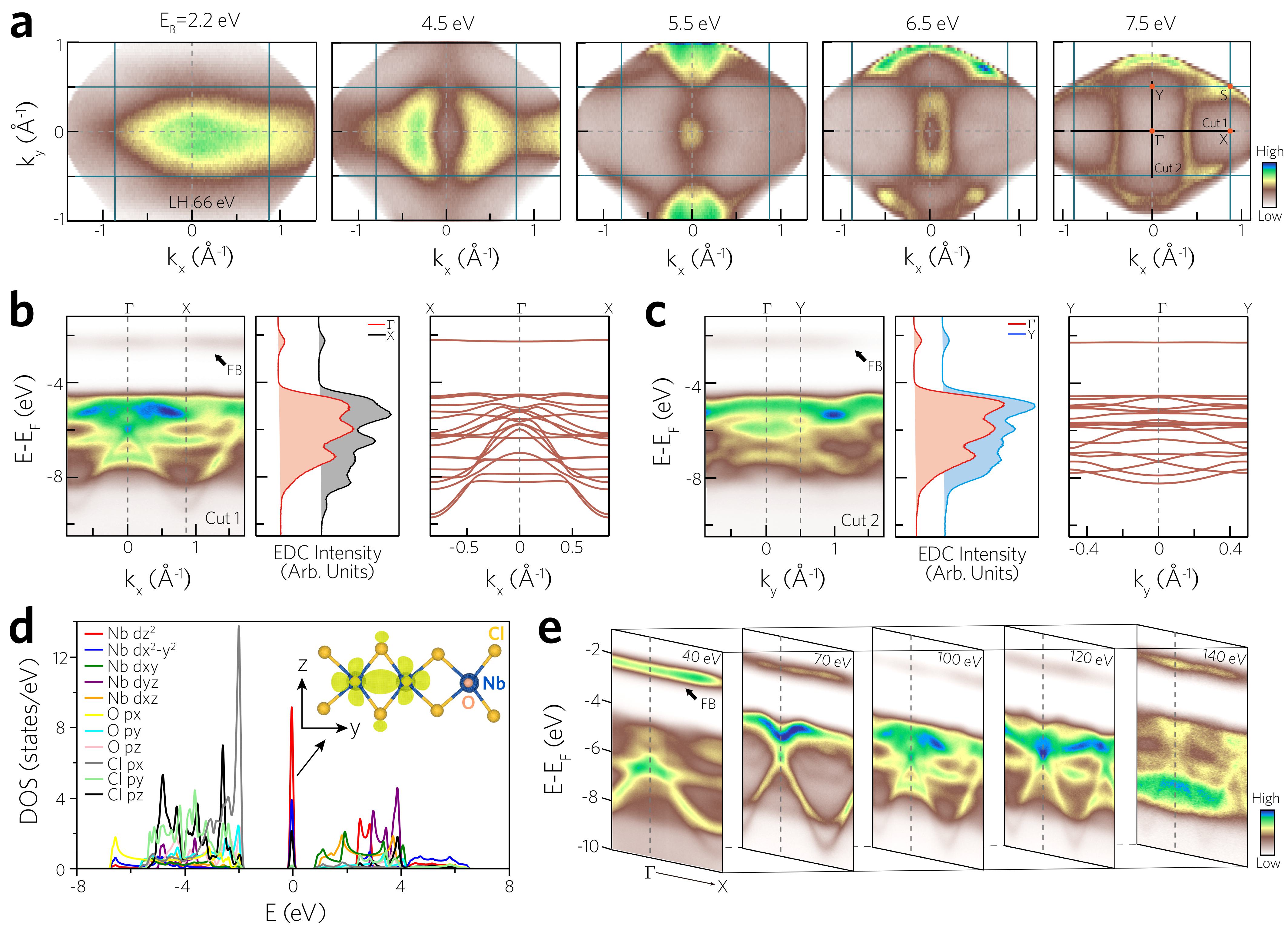}
    \end{center}
	\caption{{\bf Momentum-independent flat band and its orbital character in NbOCl$_{2}$ characterized by ARPES and DFT.}
    (a) Constant-energy contours of NbOCl$_{2}$ measured at room temperature (T= 300\,K) using photon energy h$\nu$ = 66\,eV, shown at representative binding energies $E_{B}$ = 2.2, 4.5, 5.5, 6.5 and 7.5\,eV. High-symmetry points of the BZ are marked by orange circles. Momentum cuts (Cut 1 and Cut 2) indicated by black lines are chosen for detailed analysis. (b) Band structures along the $\Gamma$–X direction (Cut 1) recorded at h$\nu$ = 100\,eV. The middle panel shows the corresponding EDCs integrated at $\Gamma$ and X points. The right panel presents the DFT band structure along the same path for comparison. (c) Same as (b), but along the $\Gamma$–Y direction (Cut 2). (d) Orbital-resolved projected density of states (PDOS) for monolayer NbOCl$_{2}$. Inset shows the real-space charge density of the FB's wavefunction from DFT. (e) Photon-energy-dependent ARPES spectra along the $\Gamma$–X direction, taken at h$\nu$= 40, 70, 100, 120 and 140\,eV, revealing the $k_{z}$ dependence of the valence bands.}
    \end{figure*}

    \begin{figure*}[tp]
    \begin{center}
    \includegraphics[width=2\columnwidth,angle=0]{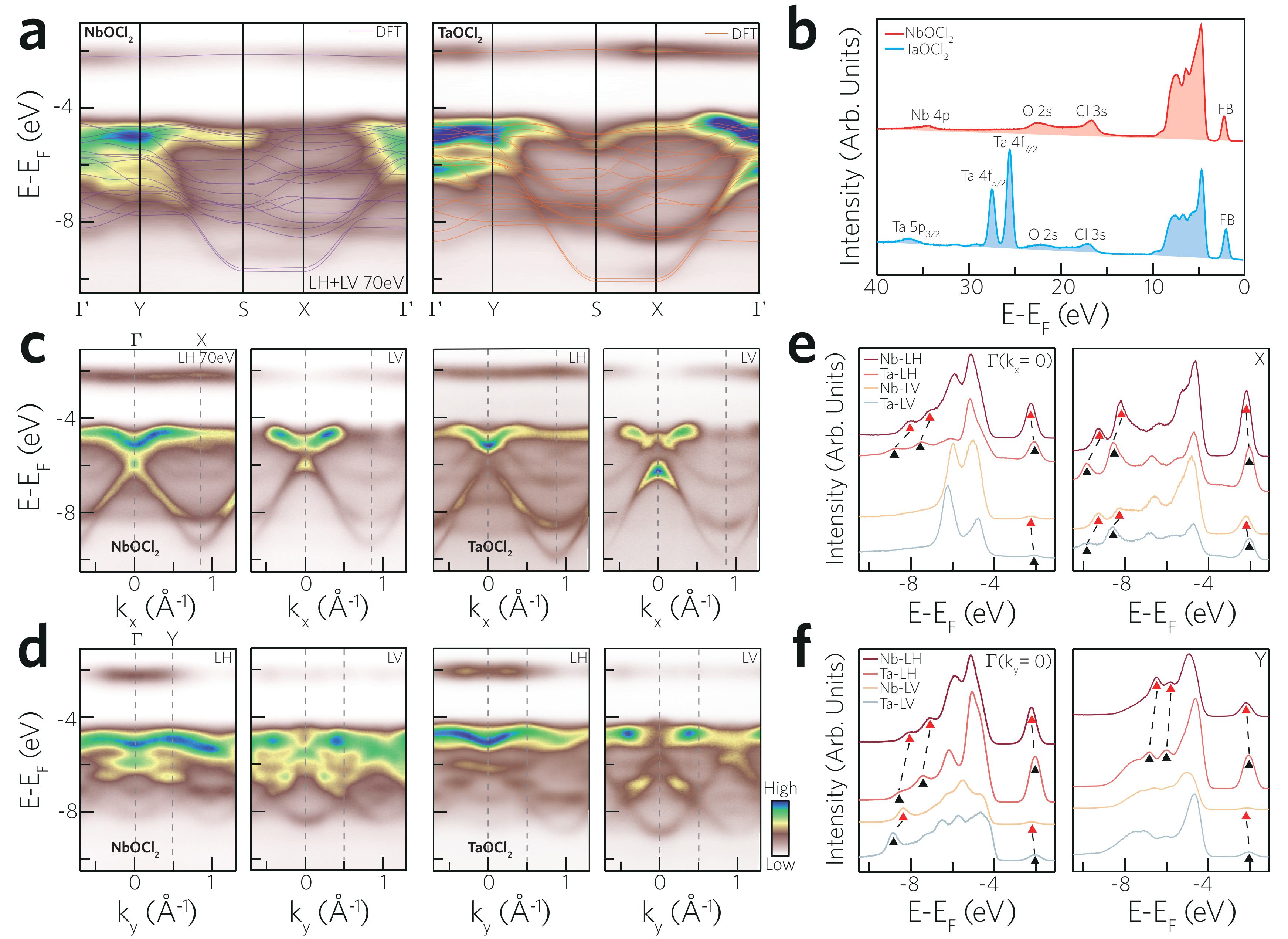}
    \end{center}
	\caption{{\bf Universality and tunability of orbital-selective flat bands across NbOCl$_{2}$ and TaOCl$_{2}$.}
    (a) Experimental band dispersions of NbOCl$_{2}$ (left) and TaOCl$_{2}$ (right) obtained by ARPES at h$\nu$= 70\,eV. Momentum cuts measured with LH and LV polarized light were combined, and are shown overlaid with DFT calculations for the corresponding monolayer structures. High-symmetry points of the BZ are defined in Fig. 1d. (b) Photoemission spectra of NbOCl$_{2}$ (red) and TaOCl$_{2}$ (blue), integrated along $\Gamma$–X direction. Core-level peaks corresponding to Nb/Ta, O and Cl elements are labeled. (c) Band dispersions along $\Gamma$–X direction for NbOCl$_{2}$ and TaOCl$_{2}$, measured separately with LH and LV light polarizations at h$\nu$ = 70\,eV. (d) Same as in (c) but along the $\Gamma$–Y direction. (e) EDCs from (c), extracted at $\Gamma$ ($k_{x}$ = 0) and X point. Peaks associated with NbOCl$_{2}$ are marked by red triangles, while those of TaOCl$_{2}$ are marked by black triangles. (f) Same as in (e) but for the data in (d).}

\end{figure*}

    \begin{figure*}[tp]
    \begin{center}
    \includegraphics[width=2\columnwidth,angle=0]{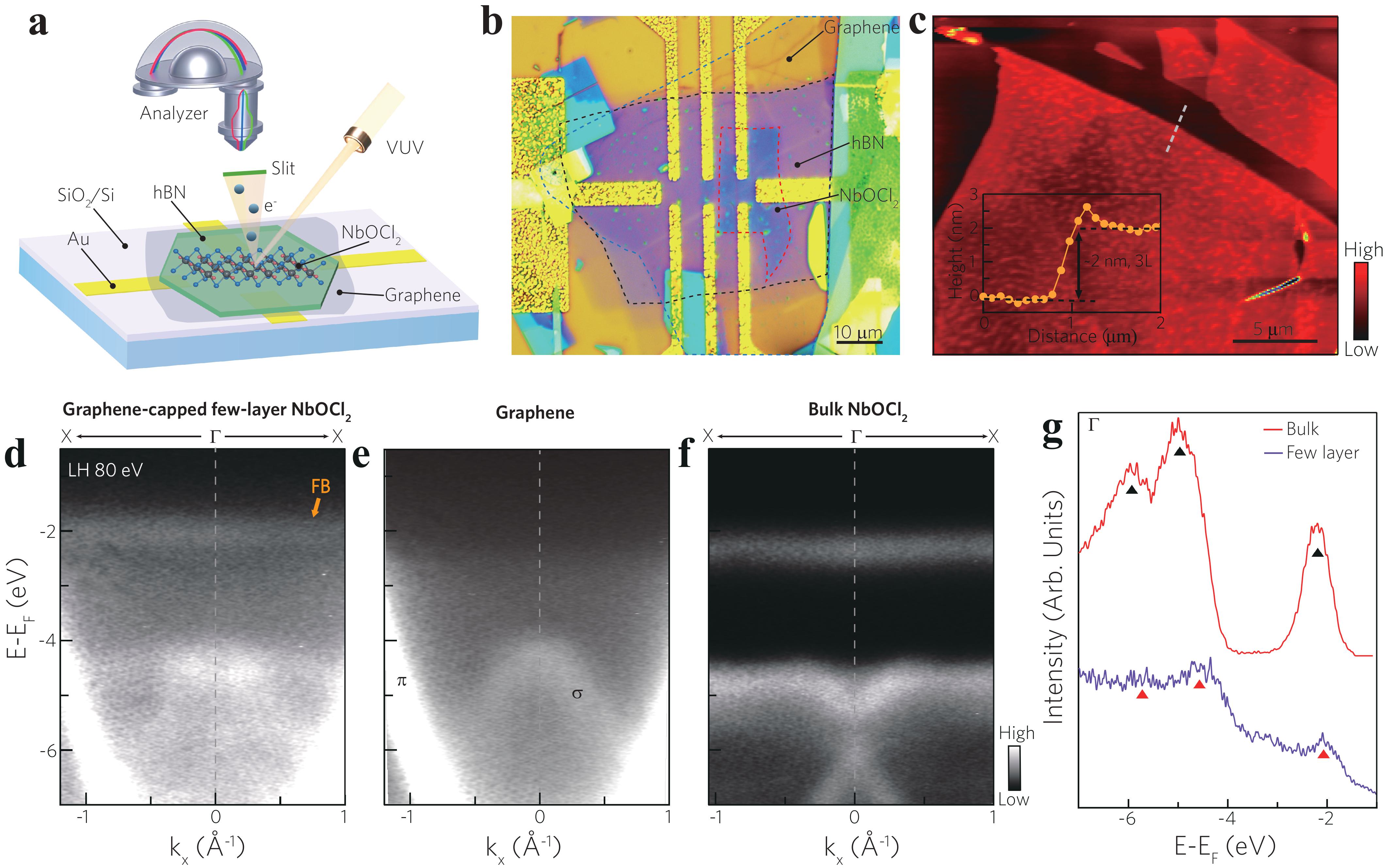}
    \end{center}
	\caption{{\bf Persistence of the flat band in few-layer NbOCl$_{2}$ revealed by ARPES.}
    (a) Schematic of the micro-ARPES setup for NbOCl$_{2}$/graphene heterostructures, with few-layer NbOCl$_{2}$ encapsulated by graphene and supported on hBN and Au contacts on a SiO$_{2}$/Si substrate. (b) Optical image of the device, showing patterned Au electrodes, hBN support, graphene encapsulation and NbOCl$_{2}$ flakes. (c) Atomic force microscopy (AFM) image of the NbOCl$_{2}$ flake with height profile (inset), confirming a thickness of $\sim$2\,nm (three layers). The position of the profile is marked by the grey dashed line. (d–f) ARPES dispersions along $\Gamma$–X for graphene capped few-layer NbOCl$_{2}$ (d), bare graphene (e) and bulk NbOCl$_{2}$ (f), measured with photon energy h$\nu$= 80\,eV under LH polarization. The flat band (FB) in the few-layer NbOCl$_{2}$ sample is marked by an orange arrow in (d). (g) EDCs at $\Gamma$ point comparing few-layer (purple) and bulk (red) NbOCl$_{2}$, showing that the flat band persists in reduced thickness.}

	\end{figure*}

    \begin{figure*}[ht]
    \begin{center}
    \includegraphics[width=2\columnwidth,angle=0]{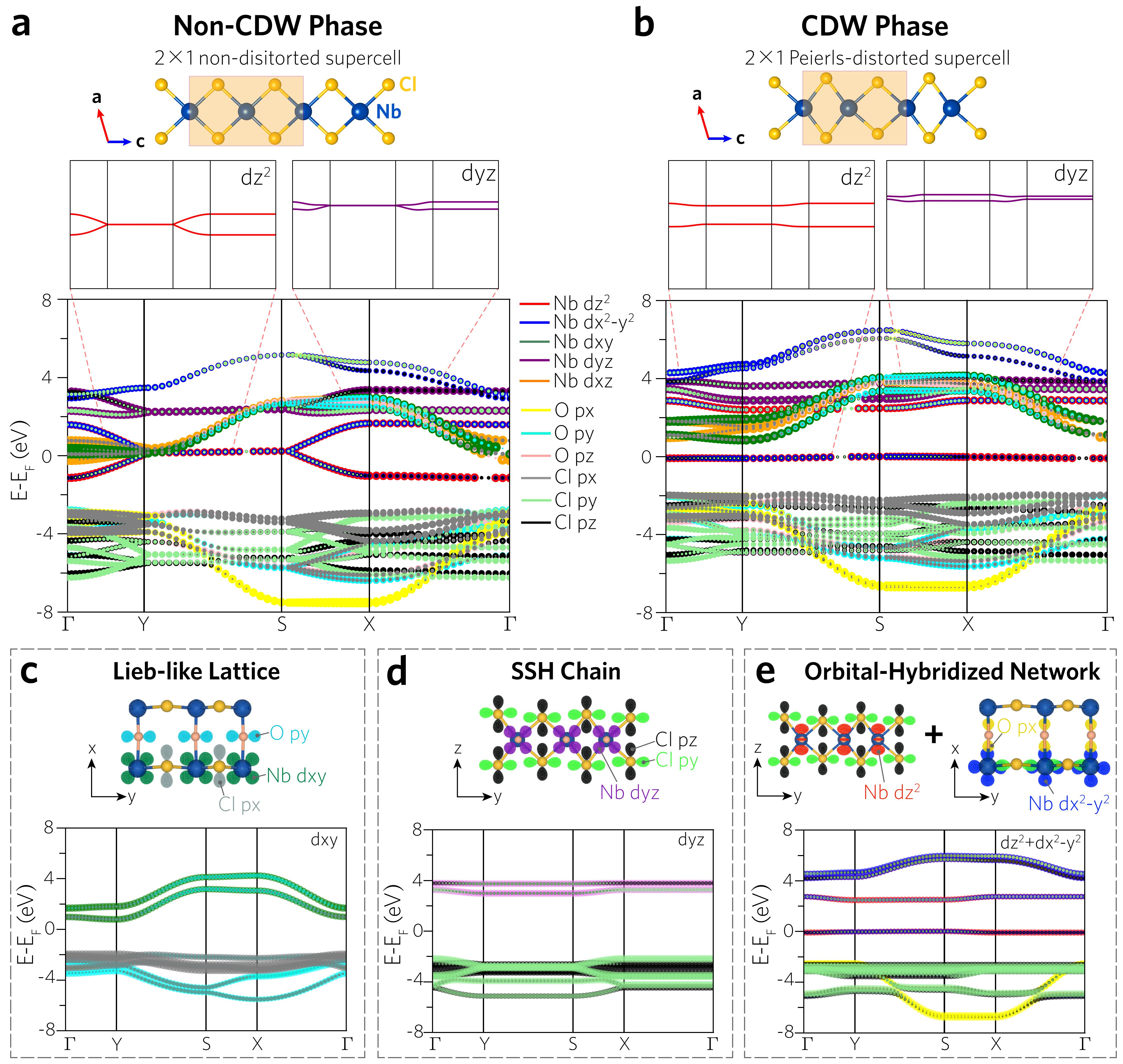}
    \end{center}
	\caption{{\bf Microscopic origin of the flat band formation from orbital hybridization and Peierls dimerization.}
    (a) Orbital-projected band structure of monolayer
    NbOCl$_{2}$ in the non-CDW phase, with the color scale indicating the relative contributions from Nb $d$ orbitals and Cl/O $p$ orbitals. The inset highlights the nearly dispersionless Nb $d_{z^{2}}$ and $d_{yz}$ bands, while retaining finite dispersion in certain momentum regions. (b) Same as in (a) but for the CDW phase. Due to the Peierls distortion, the $d_{z^{2}}$ and $d_{yz}$ -derived bands exhibit enhanced flatness relative to the non-CDW phase and open up a CDW gap, indicating an origin predominantly rooted in orbital and charge character rather than structural effects alone. (c-e) Schematic orbital–lattice building blocks and corresponding tight-binding band structures for representative real-space configurations in the CDW phase. The Lieb-lattice type configuration (c) is primarily contributed by Nb $d_{xy}$, Cl $p_{x}$ and O $p_{y}$ orbitals. The SSH chain (d) originates from Nb $d_{yz}$, Cl $p_{y}$ and $p_{z}$ orbitals. The hybridization configuration (e) arises from strong hybridization between Nb $d_{x^{2}-y^{2}}$ and $d_{z^{2}}$ orbitals that are further coupled to Cl/O $p$ states, enhancing the flatness of the flat band.}
    \end{figure*}
\clearpage

\end{document}